# Revising and Extending the Linear Response Theory for Statistical Mechanical Systems: Evaluating Observables as Predictors and Predictands

Valerio Lucarini[1,2,3] [v.lucarini@reading.ac.uk]
1. Department of Mathematics and Statistics, University of Reading, Reading, UK
2. Centre for the Mathematics of Planet Earth, University of Reading, Reading, UK
3. CEN, University of Hamburg, Hamburg, Germany

29/08/2018

## Abstract

Linear response theory, originally formulated in order to compute the response of near equilibrium statistical mechanical systems to small perturbations, has developed into a formidable set of tools for studying the forced behaviour of a large variety of systems – including out of equilibrium. Mathematically rigorous derivations of linear response theory have been provided for systems obeying stochastic dynamics as well as for deterministic chaotic systems. In this paper we provide a new angle on the problem, by studying under which conditions it is possible to perform predictions on the response of a given observable of a system to perturbations, using one or more other observables of the same system as predictors, and thus bypassing the need to know all the details of the acting. Thus, we break the rigid separation between forcing and response, which is key in linear response theory, and revisit the concept of causality. As a result, the surrogate Green functions one constructs for predicting the response of the observable of interest may have support that is not necessarily limited to the nonnegative time axis. This implies that not all observables are equally good as predictands when a given forcing is applied, as result of the properties of their corresponding susceptibility. In particular, just as in the case of some spectroscopic data inversion techniques, problems emerge from the presence of complex zeros. We derive general explicit formulas that, in absence of such pathologies, allow one to reconstruct the response of an observable of interest to N independent forcings by using as predictors N other observables. We provide a thorough test of the theory and of the possible pathologies by using numerical simulations of the paradigmatic Lorenz '96 model. Our results are potentially relevant for problems like the reconstruction of data from proxy signals, like in the case of paleoclimate, and, in general, the analysis of signals and feedbacks in complex systems where our knowledge on the system is limited, as in neurosciences. Our technique might also be useful for reconstructing the response to forcings of a spatially extended system in a given location looking at the response in a separate location.

1. **Introduction**

Response Theory provides a powerful array of methods for predicting how the statistical properties of a system change when one or more of its parameters are changed or when new forcings are added to the system. The usefulness of the theory comes from the fact that the response operators can be constructed by considering suitably defined statistical properties of the unperturbed system. This feature is particularly prominent in the case of systems with N degrees of freedom possessing invariant measure in the unperturbed state that is absolutely continuous with respect to the corresponding Lebesgue measure in N dimension. This is the case of deterministic equilibrium physical systems, whose evolution is governed by Hamiltonian dynamics and, consequently, the volume of the phase space is preserved, and in stochastically perturbed dynamical systems. In the case of full invariant measure, the Fluctuation-Dissipation theorem allows one to construct response operators from a suitably defined correlation functions in the unperturbed state [1,2,3].

Things are more difficult in the case of deterministic chaotic systems whose dynamics is governed by equations of motions featuring, on the average, a contraction of the phase space. In such systems, which can be thought as providing good models of non-equilibrium statistical mechanical systems, the fluctuation-dissipation theorem cannot be readily applied because there is no correspondence between forced and free fluctuations. This was clarified by Ruelle, who also showed that, regardless of such a complication, it is possible to construct a rigorous response theory for Axiom A dynamical systems [4,5,6]. The response theory has been later recovered using methods based on the transfer operator formalism [7], and extended to more general dynamical systems than Axiom A ones [8]. We remark that Axiom A dynamical systems can be considered of general physical relevance despite their rather specific mathematical properties than to the chaotic hypothesis[9], and the predictions of linear response theory have been verified numerically for systems that are not strictly Axiom A [10,11,12,13].

The lack of correspondence between forced and free fluctuations makes a careful study of the response theory extremely relevant. In the case of climate, response theory has been shown to be a very useful tool for predicting climate change [14,15] and for critically assessing geoengineering practices [16]. The prediction is performed by constructing response operators for each observable of interest (for one or more forcings) through a set of suitably planned test experiments, and then use such operators for predicting the response to a different time pattern of the same forcing. This approach is promising in the case one has full control of the system under investigation, thus acting is a sort of real or virtual lab, where forcing can be changed at pleasure, experiments can be repeated, and suitable observables can be measured with a good degree of precision. The direct construction of the response operator faces the extremely hard computational challenge of having to deal with contributions coming from the effects of the perturbations on the stable and on the unstable manifold of the unperturbed system, which have entirely different properties in terms of convergence [17].

Straightforward applications of the fluctuation-dissipation theorem in a chaotic system like the climate have had various degrees of success (see e.g. [18,19]). The problems mostly come from the fact that if we try to reconstruct the forced response from the free fluctuations we commit an error (due to the properties of the stable manifold of the system) which is hard to bound and which depends in a very nontrivial way on the specific choice of the forcing and of the observable of interest. In a recent paradigmatic analysis performed on a simplified climate model, it has been shown that whereas for a certain forcing the fluctuation-dissipation relation leads to a fairly good estimate of the response operator, for another choice of the forcing its skill is negligible. In this second circumstance, the response of the perturbed system can be interpreted as featuring climatic surprises, in the form of weather patterns that are entirely absent in the unperturbed system [20].

Other authors have instead taken the point of view of stochastic dynamics: in this case, linear response theory can be shown to hold under very weak assumptions, as the noise takes care of regularizing a lot of the pathologies that can be encountered as a result of deterministic dynamics [21]. This has been taken as starting point to further derive the validity of linear response theory for high-dimensional deterministic chaotic systems, by considering stochastic limits for macroscopic observables [22]. The latter point of view seems quite relevant in many complex systems where one, instead of looking at the details of the microscopic dynamics, is de facto interested in studying coarse grained variables and effective, coarse grained dynamics, which is stochastic as a result of the action of the hidden, subscale variables [23,24,25,26,27]. See also the rather comprehensive and informative review given in [28].

In this paper we want to explore a rather different direction, namely we would like to relate the response of different observables of the system to the applied perturbation. The goal is to understand to what extent it is possible to use one or more forced signals as surrogates for the actual external forcing, and then develop a response theory for predicting the response for an observable of interest based on such surrogates. The reason for looking at the response from this angle lies in the fact that whereas in the case of a direct problem we are usually knowledgeable of the specific properties of the system (including its evolution equations) and of the nature and spatio-temporal pattern of the applied forcing, this is not true when considering an inverse problem where we need to interpret data and have a possibly moderate control or knowledge of the time pattern of the applied forcing, but, in fact, we might access to multiple observables on top of the one we are specifically interested into.

The former situation is quite relevant in the case we are able to prepare and observe a system in a lab, and are able to operate on it by changing some of its parameters and the applied forcings. This is the classical *Gedankenexperiment* for which the usual response theory is relevant. The latter situation applies when our control and knowledge of the system is inherently limited, i.e. we can have access to information to a (possibly small) subset of the degrees of freedom of the system. Obvious examples of such conditions are found in neurosciences and in geosciences.

Let's expand on the specific case of geosciences, as it has provided the initial inspiration for this work. While, mathematically, one can often deal with a generic, well-behaved, observable, in natural sciences some observables are definitely more meaningful than other ones, as in the obvious case of energy when considering physical systems. Additionally, in the case of geosciences, where collecting observations and modelling the dynamics of systems is far from being trivial, our ability to observe and/or to suitably model some observables – possibly of great relevance – can be substantially more limited than for other observables. These observables can be quite different in terms of characteristic temporal (and possibly spatial) scales of variability, as when comparing atmospheric vs. oceanic fields. Such issues are particularly relevant when considering the problem of paleoclimatic reconstruction performed by using proxy data. In this case, one has access to time series (with a varying degree of uncertainty) of measurements of physical and chemical properties of natural recorders of climate variability and change (e.g. tree rings, sediments, air bubble inside ice cores, etc.). The goal is to find an approximate functional relationship between what we can actually measure – the proxy data - and what we are interested into –climate variables such as temperature, humidity, etc. [29].

Additionally, finding functional relationships between the response of different observable to climate change ('emergent constraints") and testing them against observations has been proposed as a way to assess the quality of model and reduce uncertainty on the climatic observables we have harder time to model accurately [30,31,32].

Moreover, our aim is to provide a basis for improving and extending the classical theory of feedbacks, which assumes instantaneous relations between the response of different climatic variables to forcings or, in the opposite limit, assumes that slow variables unidirectionally modulate the response of fast ones [33]. In some cases, the effect of an external forcing on a specific variable of interest is studied by looking at how it is mediated by the response of other variables and, in turn, on their influence of the variable of interest. Let's give an example: usually, in the climate literature it is said that, in global warming conditions, the increase of global precipitations is caused by the increase in the average atmospheric temperatures and, as a result – roughly speaking – via the Clausius-Clapeyron relation – of the increased ability of the atmosphere to retain water vapour [34]. Instead, both the increase of the atmospheric temperature and of the global precipitation are caused by the increase of $CO_2$ concentration, which is the real forcing to the system. Nonetheless, the statement makes indeed physical sense, as a result of the time scales of the processes. Another more specific example: it is generally thought that under global warming conditions one might expect stronger (but less frequent for a complex set of reasons) hurricanes mostly *as a result of* higher surface temperature of the ocean [35]. In order to test such a hypothesis – and isolating the effect of the surface temperature change on the statistical properties of hurricanes - a typical (yet now somewhat outdated) strategy relies on performing simulations where, instead of using coupled global climate models with prescribed forcing due to changes in the $CO_2$ concentration, one uses atmospheric general circulation models having as boundary conditions prescribed warmer sea surface temperature in the desired region [36].

We will try to provide a formal treatment of these problems based upon the linear response theory as discussed above, and we will try to define under which conditions the choice of an observable is suboptimal (or not so) in terms of the information it conveys. Can we distinguish between good observables, usable for predicting the response of other observables, and less useful ones? How do the time scales of the response of each observable to the applied forcing impact such property of bearing potential predictability? This entails defining, within a system, a (maybe approximate) chain of causality associated to the forcing between different observables, or, in the case of spatially extended systems, between different regions of the system.

As mentioned above, response theory has been successfully used in the context of climate problems and we hope to provide an additional contribution in this direction, even if our treatment is more general and applications can be foreseen in other areas of science as well.

The paper is organised as follows. In Section 2 we provide a rapid summary of the formalism of response theory, presenting it in the context of deterministic dynamical systems. We will always assume that the chaotic hypothesis [9], so that we can assume, *de facto*, that the system is close enough (for our purposes) to being uniformly hyperbolic. Nonetheless, all the subsequent results and considerations apply a fortiori in the case of rather general stochastic systems, where the conditions for the applicability of response theory are quite relaxed, as mentioned above. Section 3 provides the basic example detailing how one can use the response of one generic observable to forcing as a predictor of the response of another generic observable, partly bypassing the need to know all the details of the forcing. Explicit formulas are derived by using complex analysis and linear algebra, under assumptions - discussed later – which are needed for performing such operations. Section 4 provides the generalisation of what shown in Section 4 to the case where N forcings are applied to the system. In this case, one needs the knowledge of N generic observables to be able to predict the response of an observable of interest. Section 5 details the fundamental limitations of what discussed before. Not in all cases the operation is practically possible, as the functional relations one finds in the frequency space do not always lead to usable, causal Green functions for performing the predictions. Section 6 provides some examples of numerical experiments performed using the now classical Lorenz 96 model [37] to support our results. In Section 7 we present our conclusions and ideas for future investigations.

**2. Response Theory Formalism**

Let's consider a smooth autonomous chaotic continuous-time dynamical system acting on a smooth compact manifold $\mathcal{M}$ evolving from an initial condition $x_0$ at time $t = 0$. We define $x(t, x_0) = \Pi^t(x_0)$ its state at a generic time $t$, where $\Pi^t$ is the of evolution operator. Let's now consider the evolution for a time $t$ of a general observable $O(x)$. We define the Koopman operator $S^t$ operating on the observable $O$ as follows $S^t O = O \circ \Pi^t$, so that $S^t(O(x_0)) = O(\Pi^t(x_0))$ [38]. The corresponding set of differential equations can be customarily written as

$$\frac{dx(t)}{dt} = F(x(t))$$

(1)

where $F(y) = d/d\tau\, \Pi^\tau(y)|_{\tau=0}$. The evolution of measure $\rho$ driven by the dynamical system given in Eq. (1) is described by the Perron-Frobenius operator [38] $\mathcal{L}^t$, such that $\rho(x,t) = \mathcal{L}^t \rho(x, 0)$. The operator $\mathcal{L}^t$ is the adjoint of the operator $S^t$, and is defined as follows:

$$\int \rho(x,t)\, O(x) = \int \mathcal{L}^t \rho(x,0)\, O(x) = \int \rho(x,0)\, S^t O(x) = \int \rho(x,0)\, O(x(t,x))$$

(1b)

The family of operators $\{\mathcal{L}^t\}_{t\geq 0}$, forms a one-parameter semigroup, such that $\mathcal{L}^{t+s} = \mathcal{L}^t \mathcal{L}^s$ and $\mathcal{L}^0 = 1$. The same applies for the family of operators $\{S^t\}_{t\geq 0}$. The invariant measure $\rho_0$ of the dynamical system given in Eq. (1) is the eigenvector with eigenvalue 1 for all $\{\mathcal{L}^t\}_{t\geq 0}$. Assuming strong continuity and boundedness of the semigroup given by $\{\mathcal{L}^t\}_{t\geq 0}$, we can introduce the Liouville operator L, such that $\mathcal{L}^t = \exp(tL)$, which allows one to construct the partial differential equation for the measure mirroring the set of differential equations given in Eq. (1):

$$\frac{\partial}{\partial t}\rho = -\nabla \cdot (\rho F) = L\rho$$

(1c)

We now perturb the autonomous dynamics given in Eq. (1) by one specific forcing, so that the system is modified as follows:

$$\frac{dx}{dt} = F(x) + e_1(t) G_1(x)$$

(2)

where $G_1(x)$ is the forcing and $e_1(t)$ is its time modulation. Having in mind the example of climate change, one may think $G_1(x)$ as the extra radiative forcing due to the anomaly of $CO_2$ concentration, with $e(t)$ defining its time-pattern; see [10,11]. Assuming – as mentioned above - the chaotic hypothesis, response theory says that for any sufficiently smooth (e.g. $C^3$) observable $\Psi_1(x)$ the (time-dependent) change in the expectation value with respect to the unperturbed case can be written in linear approximation as

$$\delta\langle\Psi_1\rangle^{(1)}(t) = \int_{-\infty}^{\infty} d\tau\, \Gamma_{\Psi_1,G_1}(\tau) e_1(t-\tau)$$

(3a)

where $\Gamma_{\Psi_1,G_1}(\tau)$ is the (causal) Green function. $\Gamma_{\Psi_1,G_1}(\tau)$ can be written in terms of the properties of the unperturbed flow and depends explicitly on $\Psi_1, G_1$. Using Ruelle's response theory, we have:

$$\Gamma_{\Psi_1,G_1}(\tau) = \Theta(\tau)\int \rho_0(dx)\, G_1(x) \cdot \nabla \Psi_1(x(\tau)).$$

(3b)

Note that in the case of a system possessing an invariant measure that is smooth with respect to Lebesgue, so that $\rho_0(dx) = \rho_0(x)dx$, we can write:

$$\Gamma_{\Psi_1,G_1}(\tau) = \Theta(\tau)\int dx\, \rho_0(x)G_1(x)\cdot\nabla\Psi_1(x(\tau)) = -\Theta(\tau)\int dx\, \nabla[\rho_0(x)G_1(x)]\Psi_1(x(\tau))$$
$$= \Theta(\tau)\int dx\, \left(L_{G_1}\rho_0(x)\right)\Psi_1(x(\tau))$$

(3c)

which is a very general form of the fluctuation-dissipation theorem, where $L_{G_1}$, if $e_1(t) = 1$, is the perturbation to the Liouville operator on the right hand side of Eq. (1c) due to the introduction of the additional forcing given by $G_1(x)$.

Note that Eqs. (2)-(3) indicate that the time-dependent version of the Ruelle response theory can be used for practically constructing the time-dependent measure defining the pullback attractor [39,40,41,42] of the non-autonomous system given in Eq. (2). The results can be extended beyond linear approximation [3, 43] and applies for small (in principle, infinitesimal) forcings. Response theory becomes of little utility when one is near critical transitions [44,45], which can be seen as phase transitions for - in general - non-equilibrium systems.

After taking the Fourier Transform of Eq. (3a), we obtain:

$$\delta\langle\widetilde{\Psi_1}\rangle^{(1)}(\omega) = \tilde{\Gamma}_{\Psi_1,G_1}(\omega)\tilde{e}_1(\omega)$$

(4a)

where $\tilde{\Gamma}_{\Psi_1,G_1}(\omega)$ is usually referred to as susceptibility. We remark that any $\tilde{\Gamma}_{\Psi,G}(\omega)$, can, under suitable assumptions, be written in general as:

$$\tilde{\Gamma}_{\Psi,G}(\omega) = \sum_{k=1}^{\infty}\frac{\alpha_k\{\Psi,G\}}{(\omega - \sigma_k)}$$

(5)

where $\sigma_k$ are – for all choices of $\Psi, G$ - the eigenvalues of the Liouvillian operator $L$ introduced above, while the factors $\alpha_k$ are constructed by computing the projection of response operator on the corresponding eigenvectors. The constants $\pi_k = i\sigma_k$ are usually referred to as Ruelle-Pollicott poles [7,46,47]. If we consider smooth enough observables, causality implies that $\pi_k$ do not have positive imaginary component, i.e. $\Im[\sigma_k] = \Re(\pi_k) < 0\ \forall k$ [3,4,5]. As a result, the functions $\tilde{\Gamma}_{\Psi,G}(\omega)$ are analytic in the upper complex $\omega$ plane (and have a meromorphic extension on a strip including the real axis in the lower complex $\omega$ plane), which leads to the fact the functions $\tilde{\Gamma}_{\Psi,G}(\omega)$ obey Kramers-Kronig relations [48] for all choices of smooth $\Psi, G$ [4,5,6,49]. The presence of at least one Ruelle-Pollicott pole, say $\pi_1$, with real part very close to zero is associated to slow decay of correlations with time $t$ of the form $\approx \exp[\Re(\pi_1)t]$ in the unperturbed system and, as first discussed in [50], can lead to breakdown of linear response for small forcings; see also a detailed analysis in the case of finite dimensional Markov chains in [51]. As recently clarified in [44,45], such a condition can be considered in some cases as an effective flag for anticipating critical transitions.

One should note that the expression for the susceptibility given in Eq. (5) mirrors very closely the quantum expressions for the electric susceptibility for e.g. atoms or molecules. In this latter case, the summation involves all the pairs of the eigenstates of the unperturbed Hamiltonian operator, of the system, and in each term of the summation the imaginary part of the poles corresponds to the energy difference between the pair of considered eigenstates, the real part is the so-called line width of the transition (whose inverse is the life time), and the numerator is the so-called dipole strength for the considered transition [52]. The analytic properties of the electric susceptibility function can be associated to the fact that the system is energetically passive.

At practical level, it is important to underline that the formal similarities between linear response theory for equilibrium and nonequilibrium systems discussed in [46] allow in principle one to use algorithms developed for the analysis of electronic systems such as Vector Fitting (VF) [53] and the more advanced RKFIT [54] for the analysis of the response of multiple observables to perturbations for a general nonequilibrium system and, in particular for estimating the Ruelle-Pollicott resonances.

### 3. Observables as Predictors and Predictands: Causality, Locality, and Memory Effects

Now, let's consider a second observable $\Psi_2(x)$. Repeating the derivation as in Eq. (4) above, we find:

$$\delta\langle\widetilde{\Psi_2}\rangle^{(1)}(\omega) = \tilde{\Gamma}_{\Psi_2,G_1}(\omega)\tilde{e}_1(\omega)$$

(6a)

By taking the ratio of Eqs. (4a) and (6a), it is easy to derive that

$$\delta\langle\widetilde{\Psi_2}\rangle^{(1)}(\omega) = \frac{\tilde{\Gamma}_{\Psi_2,G_1}(\omega)}{\tilde{\Gamma}_{\Psi_1,G_1}(\omega)}\delta\langle\widetilde{\Psi_1}\rangle^{(1)}(\omega),$$

(6b)

where this relation holds regardless of the time pattern of the forcing $G_1$, but by just assuming its presence (i.e. $e_1(t)$ cannot be identically zero). The previous equation can now be written as

$$\delta\langle\widetilde{\Psi_2}\rangle^{(1)}(\omega) = \widetilde{H}_{2,1,G_1}(\omega)\delta\langle\widetilde{\Psi_1}\rangle^{(1)}(\omega)$$

(7)

where

$$\widetilde{H}_{2,1,G_1}(\omega) = \tilde{\Gamma}_{\Psi_2,G_1}(\omega)/\tilde{\Gamma}_{\Psi_1,G_1}(\omega) \quad (8)$$

or, in the time domain, after taking the inverse Fourier Transform, as

$$\delta\langle\Psi_2\rangle^{(1)}(t) = \int_{-\infty}^{\infty} d\tau\, H_{2,1,G_1}(\tau)\delta\langle\Psi_1\rangle^{(1)}(t-\tau) = H_{2,1,G_1}(t) * \delta\langle\Psi_1\rangle^{(1)}(t)$$

(9)

which, in general, is not local in time, but indeed linear. We refer to $\widetilde{H}_{2,1,G_1}(\omega)$ as the surrogate susceptibility and to $H_{1,2,G_1}(\tau)$ as the surrogate Green function. One can clearly exchange the role of $\Psi_1$ and $\Psi_2$ and derive

$$\delta\langle\Psi_1\rangle^{(1)}(t) = H_{1,2,G_1}(t) * \delta\langle\Psi_2\rangle^{(1)}(t)$$

(10)

where, clearly, $\widetilde{H}_{1,2,G_1}(\omega) = 1/\widetilde{H}_{2,1,G_1}(\omega)$. We underline that the surrogate Green functions $\widetilde{H}_{1,2,G_1}(\omega)$ and $\widetilde{H}_{2,1,G_1}(\omega)$ depend on the specific choice of the forcing $G_1$. Next, we find an expression of the integration kernels $H_{2,1,G_1}(t)$ and $H_{1,2,G_1}(t)$ in order to understand, e.g., whether one of the two observables can be seen as precursor of the other.

The first remark is that both functions $\widetilde{\Gamma}_{\Psi_1,G_1}(\omega)$ and $\widetilde{\Gamma}_{\Psi_2,G_1}(\omega)$ are analytic in the upper complex $\omega$ plane, and so is their ratio, unless $\widetilde{\Gamma}_{\Psi_1,G_1}(\omega)$ and $\widetilde{\Gamma}_{\Psi_2,G_1}(\omega)$ possess complex zeros located there. For the moment we assume that no complex zeros are found in the upper complex $\omega$ plane; we will provide in Sect. 5 an interpretation of the cases where such condition does not hold, and how this can affect our results.

In the rest of Section 3, we make some assumptions on the functional form of $\widetilde{H}_{1,2,G_1}(\omega)$ in order to derive explicit formulas that can be used to provide a more intuitive physical interpretation to our results. Nonetheless, the main findings of the paper do not depend on the approximations taken below.

If we neglect for the moment the essential spectrum of the operator $L$ in Eq. (1c), and using Eq. (5), we propose to approximate $\widetilde{H}_{2,1,G_1}(\omega)$ with a generic rational function of the form

$$\widetilde{H}_{2,1,G_1}(\omega) = \frac{\widetilde{P}_{2,1,G_1}(\omega)}{\widetilde{Q}_{2,1,G_1}(\omega)} = \frac{a\prod_{j=1}^{N}(\omega - \omega_j)}{b\prod_{k=1}^{M}(\omega - v_k)},$$

(11)

where we can consider arbitrarily large values of $N$ and $M$, and where all the $\omega_j$'s and $v_k$'s are – in general – distinct. Because of analyticity, all the constants $v_k$'s have negative imaginary part. Obviously, by the same token, also all the poles of $\widetilde{H}_{1,2,G_1}(\omega) = 1/\widetilde{H}_{2,1,G_1}(\omega)$ must be in the upper complex $\omega$ plane, so that all the constants $\omega_j$'s must also have negative imaginary part. Note that Eq. (11) implies that for values of $\omega$ larger than then the largest value of $\omega_j$'s and $v_k$'s, we have $\widetilde{\Gamma}_{\Psi_1,G_1}(\omega) \sim \widetilde{\Gamma}_{\Psi_2,G_1}(\omega)\omega^{N-M}$, so that $\omega^{N-M}$ must correspond to the ratio between the asymptotic behaviours of the two susceptibilities. We can then express $\widetilde{H}_{2,1,G_1}(\omega)$ as follows:

$$\widetilde{H}_{2,1,G_1}(\omega) = \frac{\widetilde{P}_{2,1,G_1}(\omega)}{\widetilde{Q}_{2,1,G_1}(\omega)} = \widetilde{S}_{2,1,G_1}(\omega) + \frac{\widetilde{R}_{2,1,G_1}(\omega)}{\widetilde{Q}_{2,1,G_1}(\omega)} = \widetilde{S}_{2,1,G_1}(\omega) + \widetilde{K}_{2,1,G_1}(\omega)$$

(12)

where we perform the standard division of polynomials, so that $\widetilde{S}_{2,1,G_1}(\omega)$ is a polynomial of order $N - M$ (if $N \geq M$) while $\widetilde{R}_{2,1,G_1}(\omega)$ is of order strictly smaller than $\widetilde{Q}_{2,1,G_1}(\omega)$. Clearly, $\widetilde{S}_{2,1,G_1}(\omega) = 0$ if $N < M$. A standard partial fraction expansion gives us the following expression for $\widetilde{K}_{2,1,G_1}(\omega)$

$$\widetilde{K}_{2,1,G_1}(\omega) = \frac{a}{b}\sum_{k=1}^{M}\frac{\alpha_k}{(\omega - v_k)}$$

(13)

where $\alpha_k = \widetilde{P}_{2,1,G_1}(v_k)\left(\frac{d\widetilde{Q}_{2,1,G_1}(x)}{dx}\bigg|_{x=v_k}\right)^{-1}$ and $\delta_{i,j}$ is the standard Kronecker's delta which is nonvanishing only when the two arguments are equal. We remark that the constant $v_k$ have, in general, nothing to do with the Ruelle-Pollicott poles $\pi_k = i\sigma_k$ defined above, because here we are constructing the properties of a constrained dynamics where one of the observables acts as surrogate forcing to the other.

We recall that the inverse Fourier transform of $(-i\omega)^p$ is $\delta^p(t)$, where $\delta^p(t)$ indicates the p$^{th}$ derivative of the Dirac's $\delta$ distribution, and we define: $\widetilde{S}_{1,2,G_1}(\omega) = \sum_{l=0}^{N-M}\gamma_l(-i\omega)^l$. We also recall that, assuming that $v_k$ has a negative imaginary part, the inverse Fourier transform of $1/(\omega - v_k)$ is $\Theta(t)\exp[-iv_k t]$, where $\Theta(t)$ is the Heaviside's distribution. We then obtain:

$$H_{2,1,G_1}(\tau) = \sum_{p=0}^{N-M}\gamma_p\delta^p(\tau) + \frac{a}{b}\Theta(\tau)\sum_{k=1}^{M}\alpha_k \exp[-iv_k\tau] = S_{2,1,G_1}(\tau) + K_{2,1,G_1}(\tau)$$

(14)

where we have separated the singular contribution $S_{2,1,G_1}(\tau)$ to the Green function from the non-singular one, given by $K_{2,1,G_1}(\tau)$. After performing the convolution following Eq. (9), we derive:

$$\delta\langle\Psi_2\rangle^{(1)}(t) = H_{2,1,G_1}(t) * \delta\langle\Psi_1\rangle^{(1)}(t) = \int_{-\infty}^{\infty} d\tau\, H_{2,1,G_1}(\tau)\delta\langle\Psi_1\rangle^{(1)}(t-\tau) =$$
$$\sum_{p=0}^{N-M}(-1)^p\gamma_p\frac{d^p\delta\langle\Psi_1\rangle^{(1)}(t)}{dt^p} + \frac{a}{b}\int_{-\infty}^{\infty}d\tau\,\Theta(\tau)\sum_{k=1}^{M}\alpha_k\exp[-iv_k\tau]\,\delta\langle\Psi_1\rangle^{(1)}(t-\tau)$$

(15)

The terms in the first summation provide a local (in time) link between $\delta\langle\Psi_1\rangle^{(1)}$ and $\delta\langle\Psi_2\rangle^{(1)}$. The remaining terms are due to the non-singular component of the Green function and provide the non-local (in time) contribution, where memory effects are relevant. We recall that, asymptotically, the decay of the memory is controlled by the pole with the smallest (in absolute value) real part. At finite time, the coefficients $\alpha_k$ contribute to determining which terms of the previous summation are dominant.

As discussed above, we have that $\widetilde{H}_{1,2,G_1}(\omega) = 1/\widetilde{H}_{2,1,G_1}(\omega)$. Therefore, if the integration Kernel $\widetilde{H}_{2,1,G_1}(\omega) = \widetilde{K}_{2,1,G_1}(\omega)$ contains purely non-local terms – e.g. if N<M discussed above, see Eqs. (12)-(13) - then we have that necessarily $\widetilde{H}_{1,2,G_1}(\omega)(\tau)$ will be, instead, of the form given in Eq. (15), with non-vanishing contributions from the Dirac's delta and its derivatives:

$$\delta\langle\Psi_1\rangle^{(1)}(t) = \sum_{p=0}^{M-N}(-1)^p\varphi_p\frac{d^p\delta\langle\Psi_2\rangle^{(1)}(t)}{dt^p} + \frac{c}{d}\int_{-\infty}^{\infty}d\tau\,\Theta(\tau)\sum_{k=1}^{N}\beta_k\exp[-i\omega_k\tau]\delta\langle\Psi_2\rangle^{(1)}(t-\tau)$$

(16)

where we have used new symbols for the coefficient in front of the various terms and have inserted the appropriate poles $\omega_k$'s in the last summation. Note that we should not expect the set of poles $\omega_k$ and $v_k$ to be identical, and this reflects into the fact that different observables retain differently the properties of the memory of the system.

In this case, there is a clear difference between the two observables $\delta\langle\Psi_2\rangle^{(1)}$ and $\delta\langle\Psi_1\rangle^{(1)}$, because local (in time) information is relevant only in one direction of the inference. Nonetheless, it is important to note that inference is possible from any pair of observables, in both directions. Furthermore, we note that, in this case also $\delta\langle\Psi_1\rangle^{(1)}(t)$ also plays the roles of a surrogate forcing, because it appears in the convolution integral. The memory terms in Eq. (16) drop out when $\tilde{P}_{2,1,G_1}(\omega)$ in Eq. (11) is a constant. This is in general realised when $\tilde{K}_{2,1,G_1}(\omega)$ has a constant numerator.

A clearer symmetry is established between $\delta\langle\Psi_1\rangle^{(1)}$ and $\delta\langle\Psi_2\rangle^{(1)}$ when N=M. In this case $\delta\langle\Psi_2\rangle^{(1)}(t)$ is given by a term proportional to $\delta\langle\Psi_1\rangle^{(1)}(t)$ plus a non-local term depending on $\delta\langle\Psi\rangle^{(1)}(t)$ at previous times, and the same reversing the applies reversing the roles of the two observables, even if the non-local rest term is clearly different in the two cases, compare Eq. (15) and Eq. (16) setting N=M (in this case $\varphi_0 = 1/\gamma_0$). In general, since the poles $\omega_k$ and $v_k$ are in general different, the time scales over which the memory acts in the two directions of prediction are in general different.

**4. Generalizing the Results for more Complex Patterns of Forcing**
We now wish to generalise the results presented above by allowing for accommodating multiple forcings in the system. The price to pay will be, as shown below, the need for considering multiple observables as predictors. Let's consider the case of a more complex pattern of perturbations added to the system:

$$\frac{dx}{dt} = F(x) + e_1(t)G_1(x) + e_2(t)G_2(x)$$

(17)

where $G_1(x)$ and $G_2(x)$ are two different vector fields modulated by two different time patterns $e_1(t)$ and $e_2(t)$. We now consider three independent observables $\Psi_1$, $\Psi_2$, and $\Psi_3$, whose change in the expectation value due to the introduction of the forcing can be written as:

$$\delta\langle\widetilde{\Psi_1}\rangle^{(1)}(\omega) = \tilde{\Gamma}_{\Psi_1,G_1}(\omega)\tilde{e}_1(\omega) + \tilde{\Gamma}_{\Psi_1,G_2}(\omega)\tilde{e}_2(\omega) \qquad (18a)$$
$$\delta\langle\widetilde{\Psi_2}\rangle^{(1)}(\omega) = \tilde{\Gamma}_{\Psi_2,G_1}(\omega)\tilde{e}_1(\omega) + \tilde{\Gamma}_{\Psi_2,G_2}(\omega)\tilde{e}_2(\omega) \qquad (18b)$$
$$\delta\langle\widetilde{\Psi_3}\rangle^{(1)}(\omega) = \tilde{\Gamma}_{\Psi_3,G_1}(\omega)\tilde{e}_1(\omega) + \tilde{\Gamma}_{\Psi_3,G_2}(\omega)\tilde{e}_2(\omega) \qquad (18c)$$

Following the same approach as before, we would like to be able to construct an effective equation for expressing the change in one of the observables as a function of the change in the other ones. Without loss of generality, we want to find a relationship

$$\delta\langle\widetilde{\Psi_3}\rangle^{(1)}(\omega) = \widetilde{H}_{3,1}(\omega)\delta\langle\widetilde{\Psi_1}\rangle^{(1)}(\omega) + \widetilde{H}_{3,2}(\omega)\delta\langle\widetilde{\Psi_2}\rangle^{(1)}(\omega) \qquad (19)$$

where we have slightly simplified the notation for matters of readability. We now substitute the right hand side of Eqs. (18a,b,c) in Eq. (19) and obtain:

$$\tilde{\Gamma}_{\Psi_3,G_1}(\omega)e_1(\omega) + \tilde{\Gamma}_{\Psi_3,G_2}(\omega)e_2(\omega) = \widetilde{H}_{3,1}(\omega)\tilde{\Gamma}_{\Psi_1,G_1}(\omega)e_1(\omega) + \widetilde{H}_{3,2}(\omega)\tilde{\Gamma}_{\Psi_2,G_1}(\omega)e_1(\omega) + $$
$$+\widetilde{H}_{3,1}(\omega)\tilde{\Gamma}_{\Psi_1,G_2}(\omega)e_2(\omega) + \widetilde{H}_{3,2}(\omega)\tilde{\Gamma}_{\Psi_2,G_2}(\omega)e_2(\omega)$$
$$(20)$$

Since $e_1(\omega)$ and $e_2(\omega)$ are arbitrary, we derive:

$$\begin{pmatrix} \tilde{\Gamma}_{\Psi_1,G_1}(\omega) & \tilde{\Gamma}_{\Psi_2,G_1}(\omega) \\ \tilde{\Gamma}_{\Psi_1,G_2}(\omega) & \tilde{\Gamma}_{\Psi_2,G_2}(\omega) \end{pmatrix} \begin{pmatrix} \widetilde{H}_{3,1}(\omega) \\ \widetilde{H}_{3,2}(\omega) \end{pmatrix} = \begin{pmatrix} \tilde{\Gamma}_{\Psi_3,G_1}(\omega) \\ \tilde{\Gamma}_{\Psi_3,G_2}(\omega) \end{pmatrix} \qquad (21)$$

which gives:

$$\begin{pmatrix} \widetilde{H}_{3,1}(\omega) \\ \widetilde{H}_{3,2}(\omega) \end{pmatrix} = \begin{pmatrix} \tilde{\Gamma}_{\Psi_1,G_1}(\omega) & \tilde{\Gamma}_{\Psi_2,G_1}(\omega) \\ \tilde{\Gamma}_{\Psi_1,G_2}(\omega) & \tilde{\Gamma}_{\Psi_2,G_2}(\omega) \end{pmatrix}^{-1} \begin{pmatrix} \tilde{\Gamma}_{\Psi_3,G_1}(\omega) \\ \tilde{\Gamma}_{\Psi_3,G_2}(\omega) \end{pmatrix}$$

$$\begin{pmatrix} \widetilde{H}_{3,1}(\omega) \\ \widetilde{H}_{3,2}(\omega) \end{pmatrix} = \frac{1}{\tilde{\Gamma}_{\Psi_1,G_1}(\omega)\tilde{\Gamma}_{\Psi_2,G_2}(\omega) - \tilde{\Gamma}_{\Psi_2,G_1}(\omega)\tilde{\Gamma}_{\Psi_1,G_2}(\omega)} \begin{pmatrix} \tilde{\Gamma}_{\Psi_2,G_2}(\omega) & -\tilde{\Gamma}_{\Psi_1,G_2}(\omega) \\ -\tilde{\Gamma}_{\Psi_2,G_1}(\omega) & \tilde{\Gamma}_{\Psi_2,G_1}(\omega) \end{pmatrix} \begin{pmatrix} \tilde{\Gamma}_{\Psi_3,G_1}(\omega) \\ \tilde{\Gamma}_{\Psi_3,G_2}(\omega) \end{pmatrix}$$
$$(22)$$

As all the susceptibility functions are analytic in the upper complex a $\omega$ plane, if one excludes complex zeros in the upper complex a $\omega$ plane for the denominator in the fist factor on the right hand side of Eq. (22), one derives that also $H_{3,1}(\omega)$ and $H_{3,2}(\omega)$ are also analytic in the same region. We then derive the following integro-differential equation describing the time evolution of $\delta\langle\Psi_3\rangle^{(1)}(t)$ as a function of the past and present state of $\delta\langle\Psi_2\rangle^{(1)}(t)$ and $\delta\langle\Psi_1\rangle^{(1)}(t)$:

$$\delta\langle\Psi_3\rangle^{(1)}(t) = H_{3,1}(t) * \delta\langle\Psi_1\rangle^{(1)}(t) + H_{3,2}(t) * \delta\langle\Psi_2\rangle^{(1)}(t) \qquad (23)$$

where $H_{i,j}(t) = S_{i,j}(t) + K_{i,j}(t)$, with $S_{i,j}(t)$ indicating the singular component of the integration kernel and $K_{i,j}(t)$ the non-singular one. Equation (23), similarly to Eq. (10), has the remarkable feature of applying regardless of the time patterns $e_1(t)$ and $e_2(t)$, which may be hard to access in many applications. We remark that, instead, $\widetilde{H}_{3,1}(\omega)$ and $\widetilde{H}_{3,2}(\omega)$ do depend on the vector flows $G_1(x)$ and $G_2(x)$.

The problem can be generalized to the case when N distinct forcings are applied:

$$\frac{dx}{dt} = F(x) + \sum_{l=1}^{N} e_j(t)G_j(x)$$
$$(24)$$

and we wish to express the response of one observable $\delta\langle\Psi_{N+1}\rangle^{(1)}$ as a linear combination of the response of other N observables as follows:

$$\delta\langle\widetilde{\Psi_{N+1}}\rangle^{(1)}(\omega) = \sum_{l=1}^{N} \widetilde{H}_{N+1,l}(\omega)\delta\langle\widetilde{\Psi_l}\rangle^{(1)}(\omega)$$

(25)

By expressing $\delta\langle\widetilde{\Psi_k}\rangle^{(1)}(\omega) = \sum_{l=1}^{K} \tilde{\Gamma}_{\Psi_k,G_l}(\omega)\tilde{e}_l(\omega)$ for k=1,…,N+1, where $\tilde{\Gamma}_{\Psi_k,G_l}(\omega)$ is the susceptibility of the observable $\Psi_k$ subjected to the forcing $G_l(x)$, and inserting this in equation (24), we derive by equating all terms with the same factor $\tilde{e}_l(\omega)$, l=1,…,N:

$$\begin{pmatrix} \tilde{\Gamma}_{\Psi_1,G_1}(\omega) & \cdots & \tilde{\Gamma}_{\Psi_N,G_1}(\omega) \\ \vdots & \ddots & \vdots \\ \tilde{\Gamma}_{\Psi_1,G_N}(\omega) & \cdots & \tilde{\Gamma}_{\Psi_N,G_N}(\omega) \end{pmatrix} \begin{pmatrix} \widetilde{H}_{N+1,1}(\omega) \\ \vdots \\ \widetilde{H}_{N+1,N}(\omega) \end{pmatrix} = \begin{pmatrix} \tilde{\Gamma}_{\Psi_{N+1},G_1}(\omega) \\ \vdots \\ \tilde{\Gamma}_{\Psi_{N+1},G_N}(\omega) \end{pmatrix}$$

(26)

which gives as final solution:

$$\begin{pmatrix} \widetilde{H}_{N+1,1}(\omega) \\ \vdots \\ \widetilde{H}_{N+1,N}(\omega) \end{pmatrix} = \begin{pmatrix} \tilde{\Gamma}_{\Psi_1,G_1}(\omega) & \cdots & \tilde{\Gamma}_{\Psi_N,G_1}(\omega) \\ \vdots & \ddots & \vdots \\ \tilde{\Gamma}_{\Psi_1,G_N}(\omega) & \cdots & \tilde{\Gamma}_{\Psi_N,G_N}(\omega) \end{pmatrix}^{-1} \begin{pmatrix} \tilde{\Gamma}_{\Psi_{N+1},G_1}(\omega) \\ \vdots \\ \tilde{\Gamma}_{\Psi_{N+1},G_N}(\omega) \end{pmatrix}$$

(27)

which define the N surrogate susceptibilities of the system. By applying the inverse Fourier transform to the $\widetilde{H}_{N+1,k}(\omega)$ functions, we can generalise Eq. (23) as follows:

$$\delta\langle\Psi_{N+1}\rangle^{(1)}(t) = \sum_{k=1}^{N} H_{N+1,k}(t) * \delta\langle\Psi_k\rangle^{(1)}(t)$$

(28)

Equations (10), (23), and (28) provide, in increasing level of generality, a comprehensive way for reconstructing the response of one observable of the system when one knows the response of other observables. As of the derivation above, the theory seem extremely flexible, as, in principle, no constraints exist in the choice of the predictor(s) and of the predictand. An important – in fact, essential - caveat is reported below.

**5. Remarks on the effects of the presence of complex zeros in the effective susceptibility functions**

Let's give a new, critical look at Eqs. (8)-(10). Assume now that $\tilde{\Gamma}_{\Psi_2,G_1}(\omega)$ possesses a complex zero in the upper complex $\omega$ plane. The occurrence of complex zeros is a very non-trivial property of complex functions, see [55], and has extremely relevant impacts in standard optical retrieval techniques [48]. This amounts to saying that $\Psi_2$ can have a zero response to specific bounded, physically realisable time patterns $e_1(t)$ for the vector field $G_1(x)$, with the ensuing result that some information on response of the system is lost if one looks at it through the lens of $\delta \langle \Psi_2 \rangle^{(1)}(t)$. As a result the surrogate susceptibility $\widetilde{H}_{2,1,G_1}(\omega)$ is not analytic in the same domain. This marks a clear departure from the standard case investigated in response theory.

As a result, the surrogate Green function $H_{2,1,G_1}(\tau)$ includes also non-causal contributions, so that its support is not limited to the non-negative subset of the real axis. This implies that in this case the knowledge of the past value of $\delta \langle \Psi_1 \rangle^{(1)}(t)$ is not sufficient for predicting the future values of $\delta \langle \Psi_2 \rangle^{(1)}$, or, in other terms $\delta \langle \Psi_1 \rangle^{(1)}(t)$ is an imperfect predictor of the predictand $\delta \langle \Psi_2 \rangle^{(1)}(t)$. One can say that the reduced system we are studying does not have a purely passive nature, but is instead also active, as a result of unstable feedbacks. Whereas response theory applies for all sufficiently well-behaved observables and cannot rule out the existence of susceptibility functions featuring complex zeros in the upper complex $\omega$ plane, the procedure outlined here, given the presence of a specific forcing with spatial patterns $G_1(x)$, allows one to discriminate between observables featuring or not predictive power on the future state of other observables.

Any prediction must be based only on the past data of the predictor. Therefore, in the case $H_{2,1,G_1}(\tau)$ includes non-causal components, we perform predictions using the modified surrogate Green function:

$$H'_{2,1,G_1}(\tau) = S_{2,1,G_1}(\tau) + K'_{2,1,G_1}(\tau) \qquad (29)$$

where $K'_{2,1,G_1}(\tau) = \Theta(\tau) K_{2,1,G_1}(\tau)$. Note that the Heaviside distribution takes care of enforcing causality[1]. In the case we are dealing with a problem where we need to reconstruct a signal (as in the case of proxy data in geosciences) rather than predict it, this problem is less relevant as the non-causal component of the surrogate Green function can also be used.

The same problems described here for the simple case of one forcing and one observable could emerge, a fortiori, in the case of N forcings and N observables, because the algebraic operations involved in constructing the inverse of the matrix in Eq. (30) can cause the emergence of poles in the upper complex $\omega$ plane for one or more of the surrogate susceptibilities $\widetilde{K}_{k,1}(\omega)$. As a result, we have to replace in any actual prediction the integral kernel $K_{k,1}(\tau)$ with its causal component $K'_{k,1}(\tau)$, as described above. We will show below a case where, instead, using two observables as predictors (thus taking advantage of Eq. 23) cures the pathology associated with the presence of complex zeros in the susceptibility of one of such observables.

The fact that not all choices of predictors and predictands are equally fortunate, makes, in fact, perfect intuitive sense, as implied in the classical concept of feedback on a system. What we show above is a mathematical characterization of such an intuition: when suitable predictors are chosen for predictand, the non-causal component of the integral kernel will be as small as possible (or even identically zero).

---

[1] Note that $\widetilde{K'}_{2,1,G_1}(\omega) = \widetilde{K}_{2,1,G_1}(\omega) * \left(-iP\left(\frac{1}{\omega}\right) + \pi\delta(\omega)\right)$, where P indicates the principal part..

## 6. Numerical Experiments

In order to test the performance and explore potential pitfalls of the methodology proposed in this paper, we consider the now classic and widely used Lorenz 96 model [37]. This is a forced and dissipative model that represents metaphorically the main processes – advection, dissipation, and forcing – occurring in a latitudinal circle of the atmosphere. The model has rapidly become a standard testbed for many mathematical methods in geophysics [56,57,58,59], and most notably for data assimilation and parametrization schemes, and has recently gained relevance in the context of dynamical systems theory and statistical mechanics [12,60,61,62]. In its simpler, one-level - version, the model is defined as follows:

$$\frac{dx_k}{dt} = (x_{k+1} - x_{k-2})x_{k-1} - \upsilon x_k + F_k, \quad k = 1, \dots, N$$

(30)

with periodic boundary conditions $x_{-1} = x_{N-1}$, $x_0 = x_N$, and $x_1 = x_{N+1}$. One usually assumes $\upsilon=1$ and $F_k = F, \forall k = 1, \dots, N$. The first term on the right hand side describe the nonlinear advection, the second term describes the dissipation, and the last term describes the forcing. In the inviscid and unforced case ($\upsilon = 0$ and $F = 0$), the total energy of the system $E = \sum_{k=1}^{N} x_k^2/2$ is a conserved quantity (but it is not the generator of the time translations, see [63]). When $\upsilon=1$, the model exhibits chaotic behaviour and intensive properties for $F \geq 6$ and $N \geq 20$, approximately [62]. We select for our simulations the classical values of $F = 8$ and $N = 36$, which put the system well within the chaotic regime. We perform all of our simulations below taking advantage of the MATLAB2017™ software package and using an adaptive Runge-Kutta 4$^{th}$ order time integrator with absolute and relative precision of $10^{-8}$.

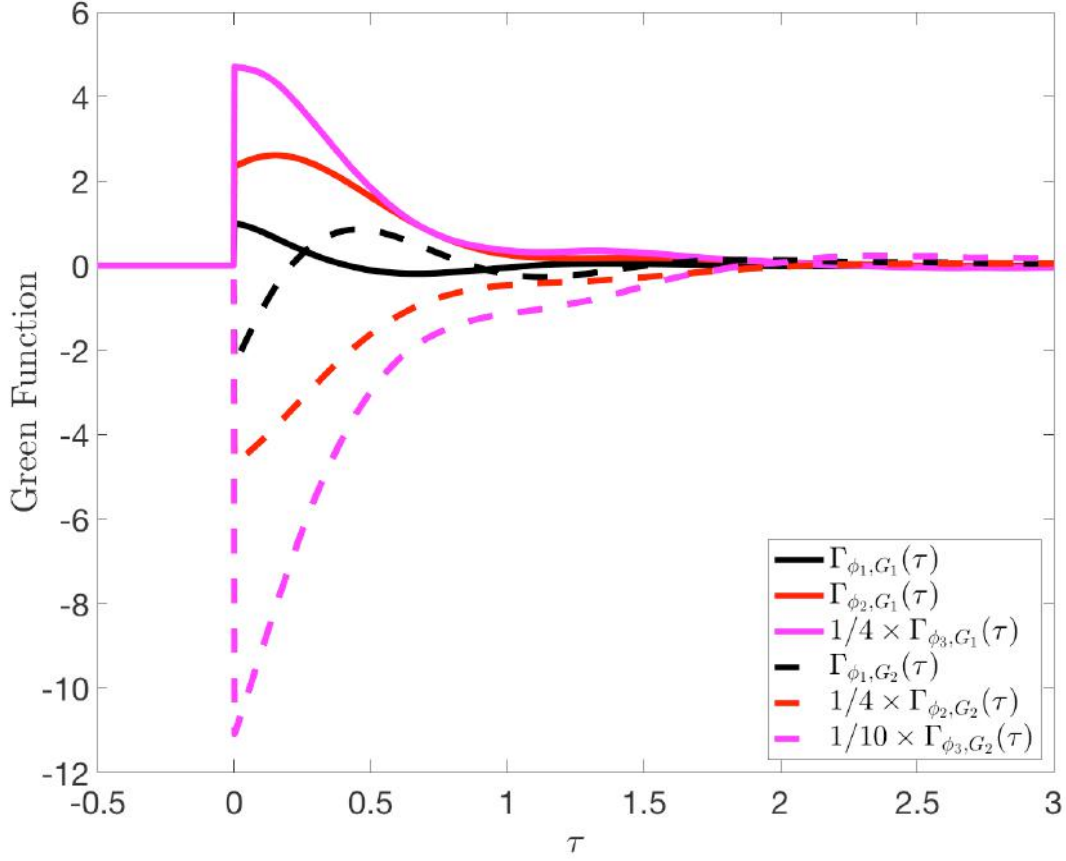

**Figure 1:** Green Functions derived for the three global observables considering perturbations to the forcing F and to the viscosity $v$ parameters.

We first run a very long simulation of the unperturbed system lasting $10^6$ time units from a random initial condition after discarding the data coming from time $10^3$ units of integration, in order to remove transient effects and be safely within the steady state regime. We then consider – separately - the following perturbation vector fields:

- $G_{1,k}(x) = 1, \forall k = 1, \ldots, N$, corresponding to a change in the constant forcing;
- $G_{2,k}(x) = -x_k \forall k = 1, \ldots, N$, corresponding to a change in the viscosity.

We investigate the response of the system in terms of changes in the expectation value of the observables $\phi_j = 1/N \sum_{k=1}^{N} x_k^j/j$, j=1, 2, 3; where, in particular, $\Psi_1$ is usually referred to as momentum of the system, and $\Psi_2$ is, as mentioned before, the energy of the system. In order to do so, we need to construct the Green functions $\Gamma_{\phi_j, G_k}(\tau)$ for all the combinations above.

Following the definition given in Eq. (3b), the Green functions need to be estimated by taking an average over the unperturbed invariant measure. We approximate it by using 100000 ensemble members, each chosen every 10 time units of the unperturbed run described above. We then proceed as follows. For each member of the ensemble, we introduce the perturbative vector field $G_1(x)$ above with the time pattern $e_1(t) = \varepsilon_1/2 \, \delta(t)$, then $e_1(t) = -\varepsilon_1/2 \, \delta(t)$, (where $\delta(t)$ is the Dirac's delta) and take the difference of the two signals for each considered observable $\phi_j$, for j=1, 2, 3. We compute the ensemble average of the signals and derive $\Gamma_{\phi_j, G_1}(\tau)$, for j=1, 2, 3. We use $\varepsilon_1 = 1$, while not small, is well within the range of linearity. Note that by computing the linear response as a centred difference, we eliminate the quadratic nonlinear term, as discussed in [20]. We then repeat the same procedure, using the same ensemble members, for the perturbative vector field $G_2(x)$, and evaluate $\Gamma_{\phi_j,G_2}(\tau)$ by computing the semidifference between the response obtained the time pattern $e_2(t) = \varepsilon_2/2\,\delta(t)$ and $e_2(t) = -\varepsilon_2/2\,\delta(t)$, We choose here $\varepsilon_2 = 0.1$, which, similarly, is well within the range of linearity.

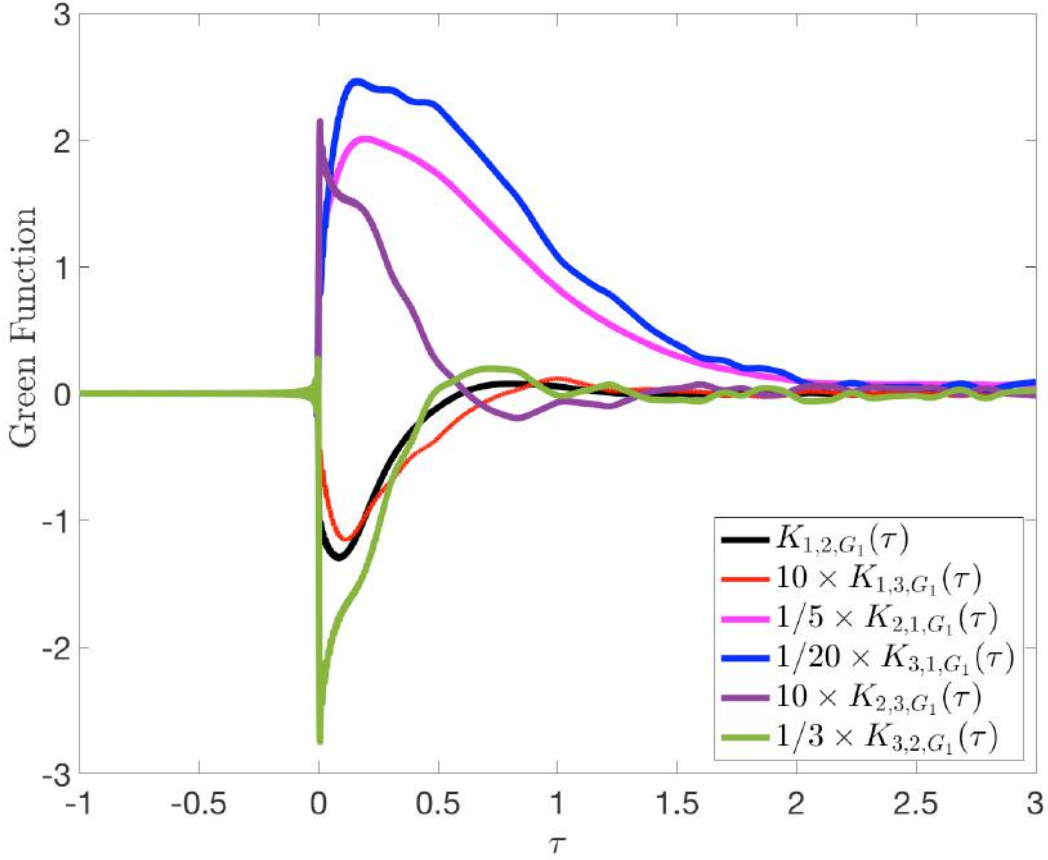

Figure 2: Non-singular components of the surrogate Green functions allowing to predict the response of one observable from the response of another observable, in the case the forcing F is perturbed.

The Green functions obtained according to this procedure are presented in Fig. 1, where we show that the characteristic response time of the perturbation to the viscosity $\upsilon$ and to the forcing F are comparable for all observables, as to be expected from Eq. (5). As quite intuitive from physical arguments and from the shape of the Green function, increasing the forcing has the effect of increasing the expectation value of the three considered observables, whereas the opposite holds for the viscosity.

All Green functions analysed here have non-vanishing values in the limit of $t$ going to zero from positive values. Along the lines of [13], by using Eq. (3c) and the definition of $\phi_j$, we can derive the following results:

$\lim_{t\to 0^+} \Gamma_{\phi_1,G_1}(t) = \int \rho_0(dx)\, G_1(x)\cdot\nabla\phi_1(x) = \frac{1}{N}\int \rho_0(dx) \sum_{k=1}^N 1 = 1$  (31a)

$\lim_{t\to 0^+} \Gamma_{\phi_j,G_1}(t) = \int \rho_0(dx)\, G_1(x)\cdot\nabla\phi_j(x) = (j-1)\int \rho_0(dx)\,\phi_{j-1}(x),\ j=2,3$  (31b)

$\lim_{t\to 0^+} \Gamma_{\phi_j,G_2}(t) = \int \rho_0(dx)\, G_2(x)\cdot\nabla\phi_j(x) = -j\int \rho_0(dx)\,\phi_j(x),\ j=1,2,3$  (31c)

Table 1: Expectation value of the three observables considered in the numerical simulations. The estimates are computed over integration lasting 100000 time units, and the uncertainties are computed as twice the standard deviation computed via Montecarlo method over 100 realizations.

|  | $j = 1$ | $j = 2$ | $j = 3$ |
|---|---|---|---|
| $\int \rho_0(dx)\, \phi_j(x)$ | $2.342 \pm 0.001$ | $9.364 \pm 0.008$ | $36.80 \pm 0.03$ |

See Table 1 for the numerical estimates of the expectation value of the observables $\phi_j, j = 1,2,3$ in the unperturbed state. Following [13], we derive that at leading order, for large values of $\omega$:

$$\tilde{\Gamma}_{\phi_k,G_l}(\omega) = i\frac{\lim_{t=0^+} \Gamma_{\phi_k,G_l}(t)}{\omega} + O(\omega^{-2})$$

(32)

We then proceed to constructing the functions $\tilde{H}_{i,j,G_k}(\omega)$ as defined in Eq. (8). As all the considered susceptibilities have the same asymptotic behaviour for large values of $\omega$, all the functions $\tilde{H}_{i,j,G_k}(\omega)$ converge asymptotically for large values of $\omega$ to a constant. One finds:

$$\lim_{\omega \to \infty} \tilde{H}_{i,j,G_k}(\omega) = \frac{\lim_{t=0^+} \Gamma_{\phi_i,G_k}(t)}{\lim_{t=0^+} \Gamma_{\phi_j,G_k}(t)} \quad (33)$$

This leads to the presence of singular contributions $S_{i,j,G_k}(\tau)$ in the corresponding integration kernels $H_{i,j,G_k}(\tau)$. The singular contributions are in the form of Dirac's deltas and have the following expression:

$$S_{i,j,G_k}(\tau) = \frac{\lim_{t=0^+} \Gamma_{\phi_i,G_k}(t)}{\lim_{t=0^+} \Gamma_{\phi_j,G_k}(t)} \delta(\tau)$$

(34)

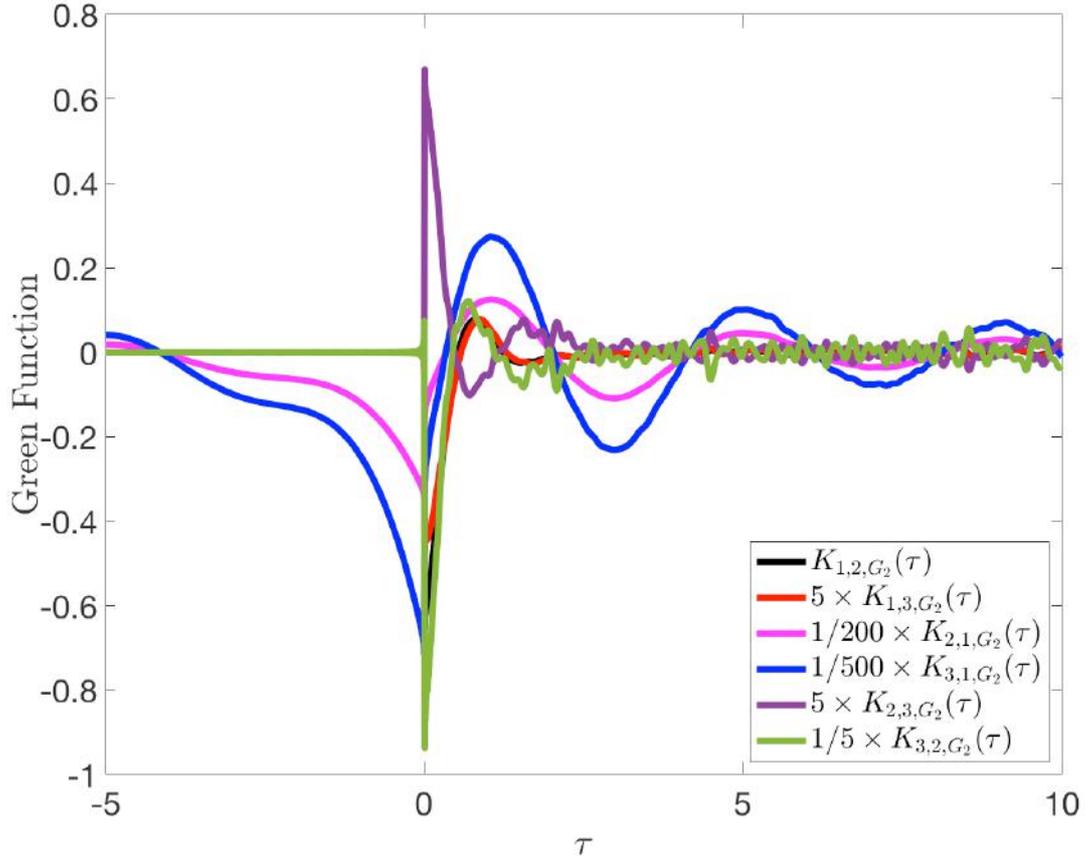

**Figure 3:** Non-singular components of the surrogate Green functions allowing to predict the response of one observable from the response of another observable, in the case the viscosity $v$ is perturbed. Note the different scale of times in the x-axis compared to the previous two figures. Note that the functions $K_{3,1,G_2}(\tau)$ and $K_{2,1,G_2}(\tau)$ have a substantial non-causal component.

We then shift our attention to the non-local component of the surrogate Green functions, for which it is harder to derive explicit results. We first treat the case of perturbations to the forcing F of the system and we show in Fig. 2 the functions $K_{i,j,G_1}(\tau)$, $i, j = 1,2,3$. The surrogate Green functions have a slightly less trivial structure than those depicted in Fig. 1, as more complex features appear, but still broadly conform to the case of representing an exponential decay modulated by oscillations. We find that in this case it is possible to reconstruct the response of each observable from the knowledge of the present (see Table 1) and past (see Fig. 2) state of any other observable. In other terms, one can treat, e.g., ɸ$_3$ as surrogate forcing for ɸ$_1$, and *viceversa*. As discussed above, the fact that, e.g., both $K_{1,3,G_1}(\tau)$ and $K_{3,1,G_1}(\tau)$ are non-vanishing proves that the neither acts as a true external forcing.

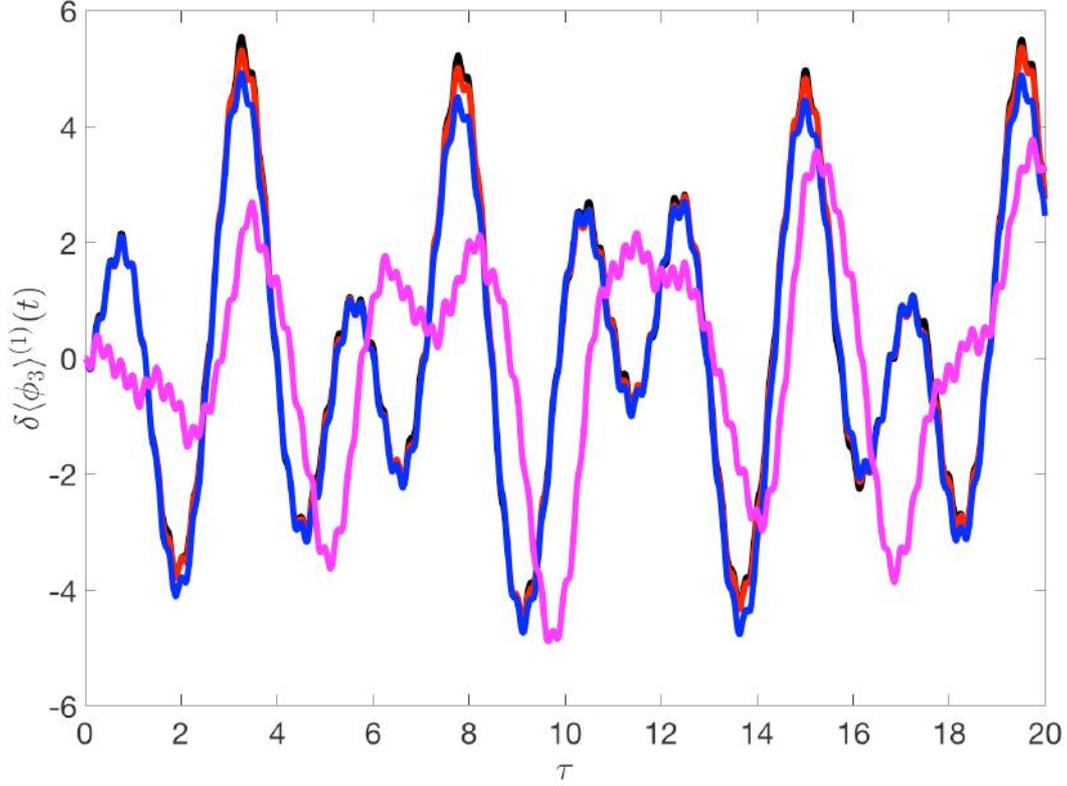

Figure 4: Predicting the response of the observable ɸ$_3$ to changes in the value of υ. Black Line: actual response of the system. Blue line: Prediction based on the surrogate Green function $H_{3,2,G_2}(\tau)$. Red line: Prediction based on classical linear response theory using the Green function $\Gamma_{3,G_2}(\tau)$, Magenta line: Prediction based on the surrogate Green function $H_{3,1,G_2}(\tau)$. Only the causal components of the Green functions are considered.

We now consider the case of perturbations to the viscosity $v$ of the system and we analyze the functions $K_{i,j,G_2}(\tau)$, i, j = 1,2,3. We find (Fig. 3) that the two functions $K_{2,1,G_2}(\tau)$ and $K_{3,1,G_2}(\tau)$ have a substantial non-causal component, and an extremely slow decay of correlations for positive times. The presence of a non-causal component *must* result from the existence of complex zeros for the function $\tilde{\Gamma}_{\Psi_1,G_2}(\omega)$ in the upper complex $\omega$ plane. Note such an occurrence is due to the specific combination of the observable and of the applied perturbation, and cannot be easily ruled out (or predicted) a priori.

We want to test the impact of the presence of a non-causal component in the surrogate Green function on our ability to predict the response of the system to a perturbation. We then consider a perturbation to the viscosity of the model (thus considering the vector field $G_{2,k}(x) = -x_k \; \forall k = 1, \dots, N$) modulated by

$$e_2(t) = 0.05 \sin(\pi t) - 0.1 \sin\left(\frac{6\pi t}{7}\right) + 0.04 \sin(8\pi t)$$

(35)

and compute over an ensemble comprising of 10000 members (the first 10% of the ensemble members described before) the time dependent change in the expectation value of ɸ$_3$, shown by the black line in Fig. 4. Clearly, the choice of $e_2(t)$ (just like $e_1(t)$ below) is arbitrary. We then estimate $\delta\langle\phi_3\rangle^{(1)}(t)$ by convolving the corresponding Green function $\Gamma_{\phi_3,G_2}(t)$ shown in Fig. 1 with $e_2(t)$ in Eq. (37). The output of the prediction of ordinary linear response theory compares quite well with the observed value and is plotted as a blue line in Fig. 4.

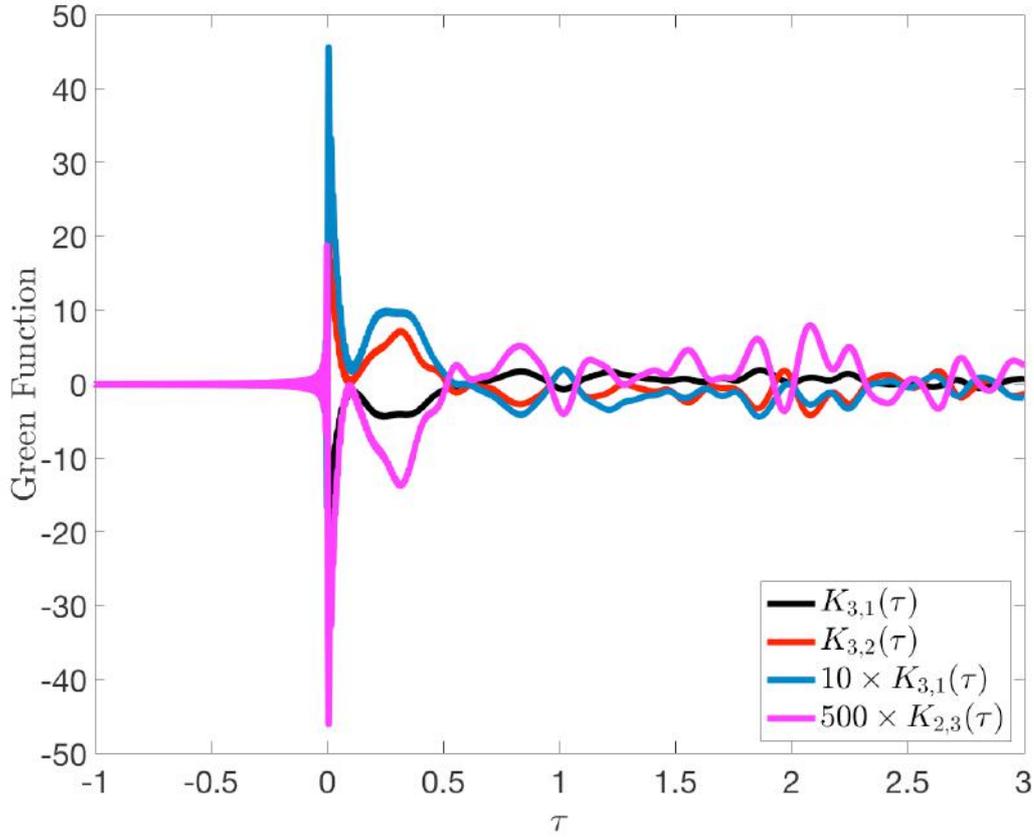

Figure 5: Non-singular components of the integration kernels associated to the matrix given in Eq. (23)

We then estimate $\delta\langle\phi_3\rangle^{(1)}(t)$ by performing the convolution of $\delta\langle\phi_2\rangle^{(1)}(t)$ with the surrogate Green function $H_{3,2,G_2}(\tau)$, whose non-singular component is causal and is shown in Fig. 3, while the singular component can be reconstructed by considering what is reported in Table 1 and Eq. (36). The prediction for $\delta\langle\phi_3\rangle^{(1)}(t)$ obtained using $\delta\langle\phi_2\rangle^{(1)}(t)$ as predictor is extremely good and is shown by the black line in Fig. 3.

Finally, we repeat the same procedure by using $\delta\langle\phi_1\rangle^{(1)}(t)$ as predictor. We then convolve $\delta\langle\phi_1\rangle^{(1)}(t)$ with the causal component of the surrogate Green function $H'_{3,1,G_2}(\tau)$, constructed as described in Eq. (29). Its Dirac's $\delta$ component can be reconstructed by looking at Table 1 and considering Eq. (36), while, for the non-singular component, the enforced causality compels us to set to zero the non-singular component $K_{3,1,G_2}(\tau)$ reported in Fig. 3 for non-positive values of $\tau$. The result of the prediction is shown as the magenta line in Fig. 4. Clearly, this latter prediction is unsuccessful, and we conclude that $\delta\langle\phi_1\rangle^{(1)}(t)$ is an inefficient predictor for $\delta\langle\phi_3\rangle^{(1)}(t)$.

We then treat the case where two independent forcings act simultaneously on the system, by adding on top of the modulation to the viscosity described in Eq. (33) a perturbation to the forcing $F$. We then consider a perturbation vector field $G_{1,k}(x) = 1 \ \forall k = 1, \dots, N)$ modulated by

$$e_1(t) = 0.125 \sin(2\pi t) - \frac{1}{6}\sin(6\pi t) + 0.2 \sin\left(\frac{10\pi t}{9}\right) \qquad (36)$$

Following Eq. (23), we aim at predicting $\delta\langle\phi_3\rangle^{(1)}(t)$ by summing the convolution of $\delta\langle\phi_2\rangle^{(1)}(t)$ with $H_{3,2}(t)$ and of $\delta\langle\phi_3\rangle^{(1)}(t)$ with $H_{3,1}(t)$. From Eq. (22) we derive the following limits:

$$\lim_{\omega\to\infty}\widetilde{H}_{3,1}(\omega) = \frac{\left[\lim_{t=0^+}\Gamma_{\phi_2,G_2}(t)\cdot\lim_{t=0^+}\Gamma_{\phi_3,G_1}(t)\right]-\left[\lim_{t=0^+}\Gamma_{\phi_1,G_2}(t)\cdot\lim_{t=0^+}\Gamma_{\phi_3,G_2}(t)\right]}{\left[\lim_{t=0^+}\Gamma_{\phi_1,G_1}(t)\cdot\lim_{t=0^+}\Gamma_{\phi_2,G_2}(t)\right]-\left[\lim_{t=0^+}\Gamma_{\phi_2,G_1}(t)\cdot\lim_{t=0^+}\Gamma_{\phi_1,G_2}(t)\right]}$$
(37a)

$$\lim_{\omega\to\infty}\widetilde{H}_{3,2}(\omega) = \frac{-\left[\lim_{t=0^+}\Gamma_{\phi_2,G_1}(t)\cdot\lim_{t=0^+}\Gamma_{\phi_3,G_1}(t)\right]-\left[\lim_{t=0^+}\Gamma_{\phi_2,G_1}(t)\cdot\lim_{t=0^+}\Gamma_{\phi_3,G_2}(t)\right]}{\left[\lim_{t=0^+}\Gamma_{\phi_1,G_1}(t)\cdot\lim_{t=0^+}\Gamma_{\phi_2,G_2}(t)\right]-\left[\lim_{t=0^+}\Gamma_{\phi_2,G_1}(t)\cdot\lim_{t=0^+}\Gamma_{\phi_1,G_2}(t)\right]}$$
(37b)

The singular components of $H_{3,1}(t)$ and $H_{3,2}(t)$ are given by $S_{3,1}(t) = \lim_{\omega\to\infty}\widetilde{H}_{3,1}(\omega)\,\delta(t)$ and $S_{3,2}(t) = \lim_{\omega\to\infty}\widetilde{H}_{3,2}(\omega)\,\delta(t)$, respectively. Instead, the non-singular components $K_{3,1}(t)$ and $K_{3,2}(t)$ are shown in Fig. 5 and are, somewhat unexpectedly, both causal. As a result, the prediction of $\delta\langle\phi_3\rangle^{(1)}(t)$ performed using Eq. (23) is virtually perfect: compare the black and the red lines in Fig. 6.

Therefore, by combining two forcings and using two observables as predictors, the pathology associated to inability of the observable $\delta\langle\phi_1\rangle^{(1)}(t)$ to be a good predictor for $\delta\langle\phi_3\rangle^{(1)}(t)$ for the specific case of forcing to the viscosity is sorted out, because more information is available on the response of the system thanks to using both $\delta\langle\phi_1\rangle^{(1)}(t)$ and $\delta\langle\phi_2\rangle^{(1)}(t)$ as predictors. In Fig. 6 we also show that $\delta\langle\phi_2\rangle^{(1)}(t)$ ends up being the most important predictor, with $\delta\langle\phi_1\rangle^{(1)}(t)$ playing, instead, a relatively minor role.

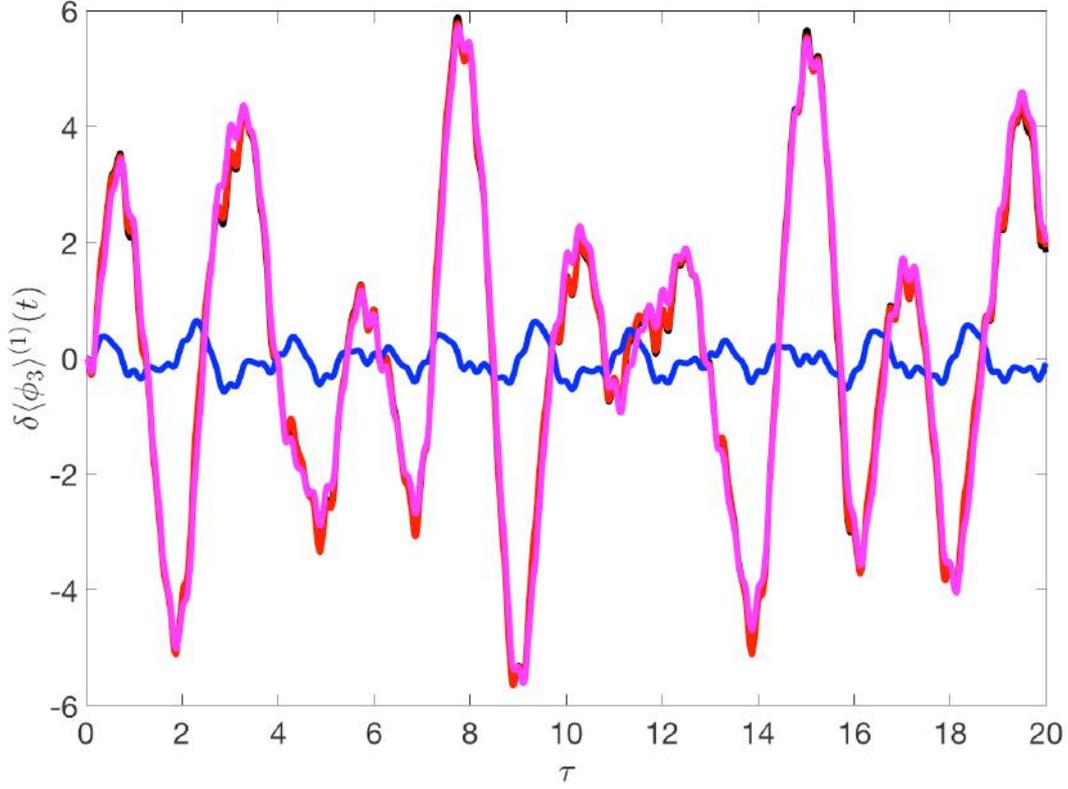

Figure 6: Predicting the response of the observable $\phi_3$ to changes in the value of F and $\upsilon$. Black Line: actual response of the system. Red line: Prediction based on the surrogate Green functions $H_{3,1}(\tau)$ and $H_{3,2}(\tau)$, using Eq. (23). Magenta line: Prediction computed by convolving only $H_{3,2}(\tau)$ with $\delta\langle\phi_2\rangle^{(1)}(t)$. Blue line: Prediction computed by convolving only $H_{3,1}(\tau)$ with $\delta\langle\phi_1\rangle^{(1)}(t)$.

A reasonable question to ask is whether Eq. (23) might be used also in the case where, in fact, only one between $e_1(t)$ or $e_2(t)$, which provide the time modulation of the perturbation vector fields $G_1(x)$ and $G_2(x)$, respectively, is different from zero. In this case, as discussed earlier, we could in principle use the simpler Eq. (10) for performing the prediction. We then decide to set $e_1(t)=0$ and use $\delta\langle\phi_1\rangle^{(1)}(t)$ and $\delta\langle\phi_2\rangle^{(1)}(t)$ as predictors of $\delta\langle\phi_3\rangle^{(1)}(t)$. Note that in Fig. 4 we had shown that in this case, using Eq. (10), $\delta\langle\phi_1\rangle^{(1)}(t)$ cannot be used to predict $\delta\langle\phi_3\rangle^{(1)}(t)$, while $\delta\langle\phi_2\rangle^{(1)}(t)$ serves the scope perfectly.

We find that, if we use Eq. (23), we can predict to a very high accuracy $\delta\langle\phi_3\rangle^{(1)}(t)$ and, more surprisingly, $\delta\langle\phi_1\rangle^{(1)}(t)$ provides a contribution to the prediction. The results are shown in Fig. 7, where the black line indicates the actual value of $\delta\langle\phi_3\rangle^{(1)}(t)$ (and matches the black line in Fig. 4), the red line shows the prediction performed using Eq. (23), while the blue line shows the contributions to the prediction coming from using $\delta\langle\phi_1\rangle^{(1)}(t)$ as predictand. Clearly, $\delta\langle\phi_2\rangle^{(1)}(t)$, which can by itself predict $\delta\langle\phi_3\rangle^{(1)}(t)$ using Eq. (10), lends some role in the predictability to $\delta\langle\phi_1\rangle^{(1)}(t)$ when the more complex Eq. (23) is used. We conjecture that using multiple predictors might increase the robustness of the prediction, as we might decrease the probability of encountering pathologies as the presence of complex zeros discussed above.

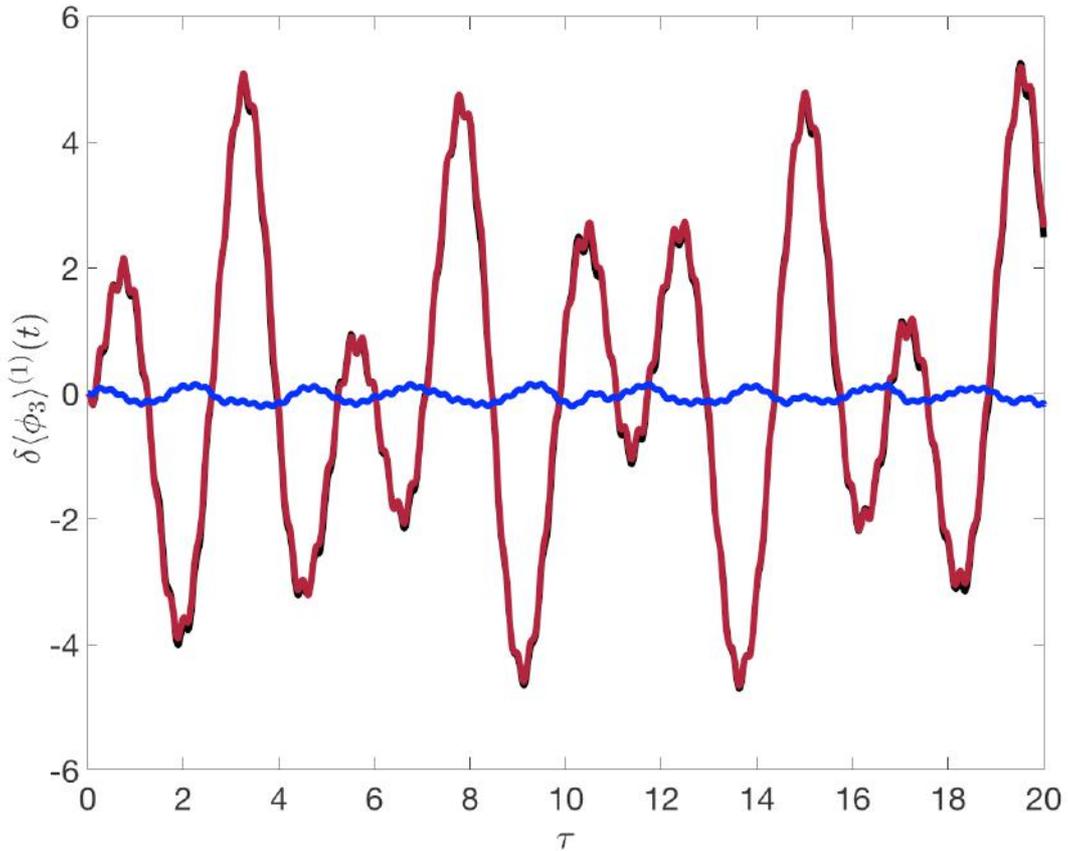

Figure 7: Predicting the response of the observable ɸ$_3$ to changes in the value of υ (as in the case depicted in Fig. 4) using two predictands $\delta\langle\phi_1\rangle^{(1)}(t)$ and $\delta\langle\phi_2\rangle^{(1)}(t)$. Black Line: actual response of the system (same as the black line in Fig. 4). Red line: Prediction based on the use of Eq. (23). Blue line: Prediction computed by convolving only $H_{3,1}(\tau)$ with $\delta\langle\phi_1\rangle^{(1)}(t)$. Note that the difference between the black curve here and that in Fig. 6 is, clearly, the blue line in Fig. 6.

**7. Summary and Conclusions**

Response theory is a well-established set of mathematical theorems or, in some cases, heuristically motivated yet often practically applicable formulas that allow one to predict how a large class of systems – including near equilibrium as well as far from equilibrium statistical mechanical systems - respond to applied perturbations. Specifically, response formulas give a prescription on how the expectation value of a generic observable will change as a result of the forcing, where such change in expressed in terms of the statistical properties of the unperturbed system. When the system has full Lebesgue measure, this boils down – in the linear case - to the celebrated fluctuation-dissipation relation. Response theory can be developed also in the case of time-dependent forcings: here, response theory provides a practical tool for approximating the time-dependent measure supported on the pullback attractor of the system under consideration. Finally, response theory – or better, the analysis of the conditions under which it breaks down – can be key for detecting and anticipating critical transitions.

While response theory is extremely powerful and widely used, it is not practically usable in some relevant scientific cases. In many complex systems we have only partial information on the state of the system, on its dynamics, on the acting forcing, and we might want to use some observables of the system as surrogate forcing. This might be motivated by practical reasons or by the knowledge of the acting feedbacks or by time scale arguments. Applications in geosciences and neuroscience often face these challenges.

In order to address this, in this paper, we have proposed a reformulation and extension of linear response theory where we blur the distinction between forcing and response, by focusing on the definition of constitutive relations relating the response of different observables, in order to construct tools for performing prediction, partly bypassing the need for knowing the applied forcing in full detail. Our theoretical predictions are supported by extensive yet extremely simple and computationally cheap numerical simulation performed on the Lorenz '96 system using commercial software (namely, MATLAB2017™) on a commercial laptop in a matter of a couple of days. The goal of these - purposely low-tech - simulations is to show that what the mathematical aspects discussed here emerge quite naturally.

We first evaluate to what extent one can use one forced signal (the change in the expectation value of one observable) to predict another forced signal (the change in the expectation value of another observable), thus bypassing the need for knowing some details (in particular, the time pattern) of the applied forcing. This leads to defining surrogate susceptibilities and surrogate Green functions, which, in practice, amount to justifying the existence of general linear relations between the various forced signals. This point of view is very natural as it reflects the way one looks at feedbacks inside a complex systems, which often entails constructing approximate temporally ordered causal links describing how the effect of the forcing propagates between different components or parts of the system itself. We explain that, in some cases, one can predict the future state of observable $\phi_2$ from the past and present knowledge of observable $\phi_1$, and vice-versa, so that it makes no sense to say what causes what. Additionally, once a forcing is considered, in agreement with physical intuition, we prove that in some cases an observable $\phi_1$ is not capable of acting as effective predictor for another generic observable $\phi_2$. This results from the presence of complex zeros in the upper complex angular frequency plane of the susceptibility of $\phi_1$. The presence of a complex zero implies that for some time-pattern of forcing, the change in the expectation value of the observable $\phi_1$ is identically zero  As a result, the surrogate Green function is not causal and  $\phi_1$ cannot be used for predicting any other observable $\phi_2$.  Such a loss of causality is obviously just apparent, because we are not relating the true forcing to a response. What we see is, in fact, the result of the loss of information due to using $\phi_1$ as surrogate for the forcing.

Finally, we show using standard linear algebra that, in absence of mathematical pathologies, as the one just mentioned above, if a system undergoes N different simultaneous forcings, it is possible to predict the response of an observable $\phi_{N+1}$ using N suitably defined surrogate Green functions convoluted with N other observables $\phi_1, ..., \phi_N$ without the need to know the time patterns of the actually applied forcings. As shown in our numerical example, such a strategy may overcome the difficulties associated to choosing a bad observable as predictor: the use of multiple predictors seems inherently more robust.

The surrogate Green functions discussed above can be derived from specific sets of experiments where only one of the forcings is applied at a time. Following [64], this can achieved also using stochastic perturbations.

This allows for a model-assisted reconstruction of signals where we have no or partial information on the time patterns of the forcings, or even if one or more of the forcings are, in fact, even inactive. This seems extremely promising for a variety of problems in the analysis of the response of complex models to perturbations where we might be interested in finding relationships between different forced signals, or when the actual forcings might be hard to control or measure.

We foresee applications of our results in problems relevant for climate science such as the Intercomparison of climate models regarding their response to external forcing, the analysis of the relationship between forced and free variability, and, on a different note, on the reconstruction of climate readings from multiple proxy data. Additionally, our results seem relevant for studying, in spatially extended systems, the response of a part of the system to perturbations by looking at its response somewhere else. This might be relevant in fluid dynamical systems, in systems obeying diffusion laws, neural systems, where the presence of a maximum speed of propagation of the information might lead to barrier to prediction if the considered parts of the system and/or the location of the forcings are too far away.

Finally, we believe that our results might have some relevance in the context of the theory of causal networks [65], because we are able to define whether or not one can assume causal relations between different observables or different regions of a system.

**Acknowledgements**

The author is indebted to P. Cox for suggesting to look into the problem of emergent constraints in climate and being patient and kind enough to wait a very long time for a project report on this topic to be prepared by the author; S. Ciliberto, T. Bodai, and A. Tantet for critical discussions on some early results; N. Boers for some inspiring conversations on the problem of inverse modelling and proxy data; A. Gritsun and M. Colangeli for many discussions on response theory; D. Ruelle for providing comments on an early version of the manuscript; T. Shepherd for suggesting links with the theory of causal networks; and to three anonymous reviewers for constructive criticisms. The paper has mostly been written during the workshop Nonequilibrium in Geosciences, Life Science, and Physics held in May 2018 at ICTP in Trieste. The author is thankful to the ICTP for proving a wonderful and creative atmosphere. This work has been partially supported by the Horizon2020 projects Blue-Action (grant No. 727852) and CRESCENDO (grant No. 641816) by the DFG SFB/Transregio project TRR181., and by the Royal Society Bilateral UK-Russia grant.